\documentclass[letterpaper,aps,prc,superscriptaddress,nofootinbib,showpacs,floatfix,reprint]{revtex4-1}
\usepackage[utf8]{inputenc}
\usepackage{amsmath}
\usepackage{amsfonts}
\usepackage{amssymb}
\usepackage{amsthm}
\usepackage{bm}
\usepackage[english]{babel}
\usepackage{fontenc}
\usepackage{graphicx}
\usepackage[margin=1in]{geometry}
\usepackage{enumerate}
\usepackage{color}

\begin{document}
\title{Investigating late-stage particle production in pp collisions with Balance Functions}
\author{ Alexandru~Manea }
\email{alexandru.manea@cern.ch}
\affiliation{Faculty of Physics, University of Bucharest, Magurele, 077125, Romania}
\affiliation{Institute of Space Science -- INFLPR Subsidiary, Magurele, 077125, Romania}
\author{ Claude~Pruneau }
\email{claude.pruneau@wayne.edu}
\affiliation{Department of Physics and Astronomy, Wayne State University, Detroit, 48201, USA}
\author{Diana~Catalina~Brandibur}
\email{diana.catalina.brandibur@cern.ch}
\affiliation{Faculty of Physics, University of Bucharest, Magurele, 077125, Romania}
\affiliation{Institute of Space Science -- INFLPR Subsidiary, Magurele, 077125, Romania}
\author{Andrea~Danu}
\email{andrea.danu@cern.ch}
\affiliation{Institute of Space Science -- INFLPR Subsidiary, Magurele, 077125, Romania}
\author{ Alexandru~F.~Dobrin }
\email{alexandru.florin.dobrin@cern.ch}
\affiliation{Institute of Space Science -- INFLPR Subsidiary, Magurele, 077125, Romania}
\author{ Victor~Gonzalez }
\email{victor.gonzalez@cern.ch}
\affiliation{Department of Physics and Astronomy, Wayne State University, Detroit, 48201, USA}
\author{ Sumit~Basu }
\email{sumit.basu@cern.ch}
\affiliation{Division of Particle and Nuclear Physics, Department of Physics, Lund University, Box 118, SE-221 00 Lund, Sweden}
\affiliation{International Center for High Energy Physics and Applications, Lovely Professional University, Punjab-144411, India}
%

\begin{abstract}
Balance functions have been regarded in the past as a method of investigating the late-stage hadronization found in the presence of a strongly-coupled medium. They are also used to constrain mechanisms of particle production in large and small collision systems. Measurements of charge balance functions for inclusive and identified particle pairs are reported as a function of charged particle multiplicity in proton--proton collisions simulated with the PYTHIA8 and the EPOS4 models. The charge balance functions of inclusive, pion, kaon, and proton pairs exhibit amplitudes and shapes that depend on particle species and differ significantly in the two models due to the different particle production mechanisms implemented in PYTHIA and EPOS. The shapes and amplitudes also evolve with multiplicity in both models. In addition, the evolution of the longitudinal rms width and that of balance functions integrals with multiplicity (and average transverse momentum) feature significant differences in the two models.
\end{abstract}

\maketitle

\section{Introduction}

Studies at the Relativistic Heavy Ion Collider (RHIC) and the Large Hadron Collider (LHC) have established that a medium akin to a quark--gluon plasma (QGP) is produced in collisions of large nuclei (e.g., Au--Au and Pb--Pb)~\cite{STAR:2005gfr,PHENIX:2004vcz,BRAHMS:2004adc,PHOBOS:2004zne,ALICE:2022wpn,CMS:2024krd}. The focus of both experimental and theoretical communities has recently largely shifted towards ``high-precision" measurements of the properties of the QGP~\cite{Arslandok:2023utm}. However, there is a keen interest in establishing whether QGP matter is also produced in high-multiplicity p--A and pp collisions and several studies have thus been conducted to measure the presence of collective flow~\cite{CMS:2016fnw, ATLAS:2017rtr, ALICE:2019zfl}, energy loss~\cite{ALICE:2023plt}, and strangeness enhancement~\cite{ALICE:2016fzo} in such collision systems. 
The results point to collective anisotropic flow and strangeness saturation in high-multiplicity pp collisions but no clear indication of jet quenching~\cite{ALICE:2023plt} or longitudinal viscous effects~\cite{ALICE:2022hor}. One can evidently argue that QGP systems formed in pp collisions might be too small or too short lived to manifest sizable/measurable jet energy-loss effects or viscous longitudinal transverse momentum $p_{\rm T}p_{\rm T}$ correlation broadening. One may then also argue that
evidence for the production of QGP matter in pp collisions is somewhat questionable
and it is consequently of interest to seek additional lines of evidence based on other observables sensitive to the presence of deconfined quark matter. One such set of observables involves charge, strangeness, and baryon number  balance functions. Measurements of charge balancing  
first appeared in the late 1970s towards the study of charged hadron production~\cite{ACCDHW:1979mjc,EHSNA22:1989cad} but were eventually formalized, as charge balance functions, at the beginning of the RHIC program as a sensitive indicator of the presence of a relatively long-lived isentropic period of hydrodynamic expansion in A--A collisions separating quark production at the onset of collisions and a short stage of delayed hadronization~\cite{Bass_2000,Jeon:2001ue}. In peripheral A--A collisions, corresponding to minimum bias pp collisions, the isentropic expansion 
stage is expected to be non-existent or rather short. However, for mid- to most-central collisions, this stage is expected to grow in some proportion of the system size and produced particle multiplicity. In a regime of isentropic expansion, quarks ($\rm q$) and anti-quarks ($\rm \overline q$) constantly annihilate and reform anew by means of gluon--gluon interactions. As the system expands longitudinally,  the $\rm q$ and $\rm \overline q$ then increasingly separate longitudinally as a result of the expansion of the system. Pairs produced at the onset of collisions have the longest time to separate and will then be found at large rapidity difference whereas pairs emitted later, e.g. just before hadronization, will be observed at much smaller rapidity separation. 
A substantial reduction of the  width of balance function between $\rm q$ and $\rm \overline{q}$ is thus a telltale indicator of a prolonged isentropic expansion stage. Although the nuclear fireballs produced in pp collisions were not initially expected to feature hydrodynamic and isentropic expansion, flow-like transverse anisotropy and strangeness saturation observed in high-multiplicity collisions~\cite{CMS:2016fnw, ATLAS:2017rtr, ALICE:2019zfl, ALICE:2016fzo} provide a compelling argument justifying the studies of the evolution of balance functions  with 
the charged particle multiplicity produced in pp collisions at both RHIC and LHC energies. In order to further substantiate this argument, it is obviously of interest to examine and contrast  predictions of models that do and do not feature a   hydrodynamic expansion stage. The task is however not entirely trivial because, meaningful simulations of balance functions of charged hadrons require the models explicitly feature (quasi)local charge conservation. Local charge conservation is evidently built-in ab-initio in a model  such as PYTHIA8~\cite{PYTHIA_manual} based on Lund string fragmentation~\cite{Andersson:1983ia} but it is considerably more difficult to implement in models featuring a hydrodynamic expansion phase. As it happens, the most recent implementation of the EPOS model, EPOS4~\cite{Werner:2023jps}, describes both small and large collision systems with a mixture of corona and core components. The corona describes particle production based on string decay whereas the core is modeled with a hydrodynamic phase. The core expansion is concluded with a particlization prescription involving local microcanonical energy-momentum and charge conservation. It thus nominally constitutes an ideal candidate to study the impact of hydrodynamic (isentropic) expansion of the fireballs produced in pp collisions based on charge balance functions and contrast these predictions with those of PYTHIA, which does not feature such expansion. The goal of this paper is thus to examine and compare predictions of charged balance functions by PYTHIA8 and EPOS4 to find whether one could meaningfully expect a finite size reduction of  the longitudinal width of balance functions in pp collisions relative to predictions of a string fragmentation model such as PYTHIA. 

This paper is organized as follows. Section~\ref{sec:method} presents the definition of the charged balance function computed in this work from PYTHIA and EPOS simulations, as well as the different kinematic and selection conditions and technical details of the computations. The balance functions of inclusive, pion, kaon, and proton pairs obtained from both models are reported as a function of charged particle multiplicity in Sec.~\ref{sec:results}. A summary of this work and conclusions are presented in Sec.~\ref{sec:summary}.

\section{Observable Definition and Analysis Method}
\label{sec:method}

The notion of balance function was initially introduced based on differences of conditional number densities of unlike-sign and like-sign pairs expressed as function of the rapidity and azimuthal angle differences between hadrons of interests~\cite{Bass:2000az}. However, balance function studies based on simulations of pp and p$\rm\overline{p}$ collisions, computed with PYTHIA, showed that integrals of the  original BF definition do not properly converge to unity when integrated over full rapidity and transverse momentum acceptance~\cite{Pruneau:2022brh}. Sum rules are recovered by replacing conditional densities by normalized two-particle differential cumulants~\cite{Pruneau:2023zhl}. In this work, balance functions are then computed according to 
\begin{widetext}
\begin{equation}
\label{eq:BF}
    B^{\alpha \bar{\beta}}(\Delta \eta, \Delta \varphi) = \bar\rho_1^{\bar{\beta}}  \left[ R_2^{\alpha\bar{\beta}}(\Delta \eta, \Delta \varphi) - R_2^{\bar{\alpha}\bar{\beta}}(\Delta \eta, \Delta \varphi) \right],
\end{equation}
\end{widetext}
where
\begin{equation}
\label{eq:R2}
    R_2^{\alpha\beta}(\Delta \eta, \Delta \varphi) = \frac{\rho_2^{\alpha\beta}(\Delta \eta, \Delta \varphi)}{\rho_1^{\alpha}\otimes \rho_1^{\beta}(\Delta \eta, \Delta \varphi)} - 1.
\end{equation}
The labels $\alpha, \beta$ denote the particle species of interest, whereas $\bar\alpha$ and $\bar\beta$
denote their respective  antiparticles.
The notation $\rho_1^{\beta}(\eta,\varphi)$ represents the single particle yield of species $\beta$ at pseudorapidity $\eta$ and azimuth $\varphi$, and $\rho_2^{\alpha\beta}(\Delta \eta, \Delta \varphi)$ is the pair density of particles of species $\alpha,\beta$ evaluated at relative pseudorapidity $\Delta \eta=\eta_1-\eta_2$ and relative azimuthal angle $\Delta\varphi= \varphi_1-\varphi_2$. 
However, whenever identified particles are considered, their  rapidity $y$ is used in the computation of densities and correlation functions rather than their pseudorapidity.
The notation  
$\rho_1^{\alpha}\otimes \rho_1^{\beta}(\Delta \eta, \Delta \varphi)$ represents the ``tensor product" of the single particle densities averaged at $\Delta \eta, \Delta \varphi$ according to a technique described in Ref.~\cite{Adam:2017ucq}. Additionally, the notation $\bar\rho_1^{\bar{\beta}}$ refers to the average of $\rho_1^{\bar{\beta}}(\eta,\varphi)$ across the acceptance of the measurement or calculation. 

Calculations of the balance functions presented in this work were carried out based on pp collisions at $\sqrt{s}=13.6$ TeV simulated with the PYTHIA8~\cite{PYTHIA_manual} and the EPOS4~\cite{Werner:2023jps} event generators. In total, $1.5 \times 10^{9}$ and $1.5 \times 10^{8}$ minimum-bias events were generated with the PYTHIA8 and EPOS4 generators, respectively, and were analysed at the Institute of Space Science -- INFLPR Subsidiary (ISS) computing grid. The simulations based on PYTHIA8 were carried out with the MONASH 2013 tune~\cite{Skands:2014pea} with soft QCD processes and color reconnection. A model based on ``core-corona" decomposition coupled to a hadronic afterburner was used for EPOS4. The study of the evolution of balance functions with produced particle multiplicity was carried out using seven multiplicity classes defined as fractions of the event samples, based on charged particles produced in $-5.0<\eta<-3.0$ and with $0.2<p_{\rm T}<3.0$~GeV/$c$, and denoted ``0--5\%", ``5--10\%", ``10--20\%", ``20--40\%", ``40--60\%", ``60--80\%", and ``80--100\%" from the highest to the lowest multiplicity.

\begin{figure}[hbt]
    \centering
    \includegraphics[width = 0.48\textwidth]{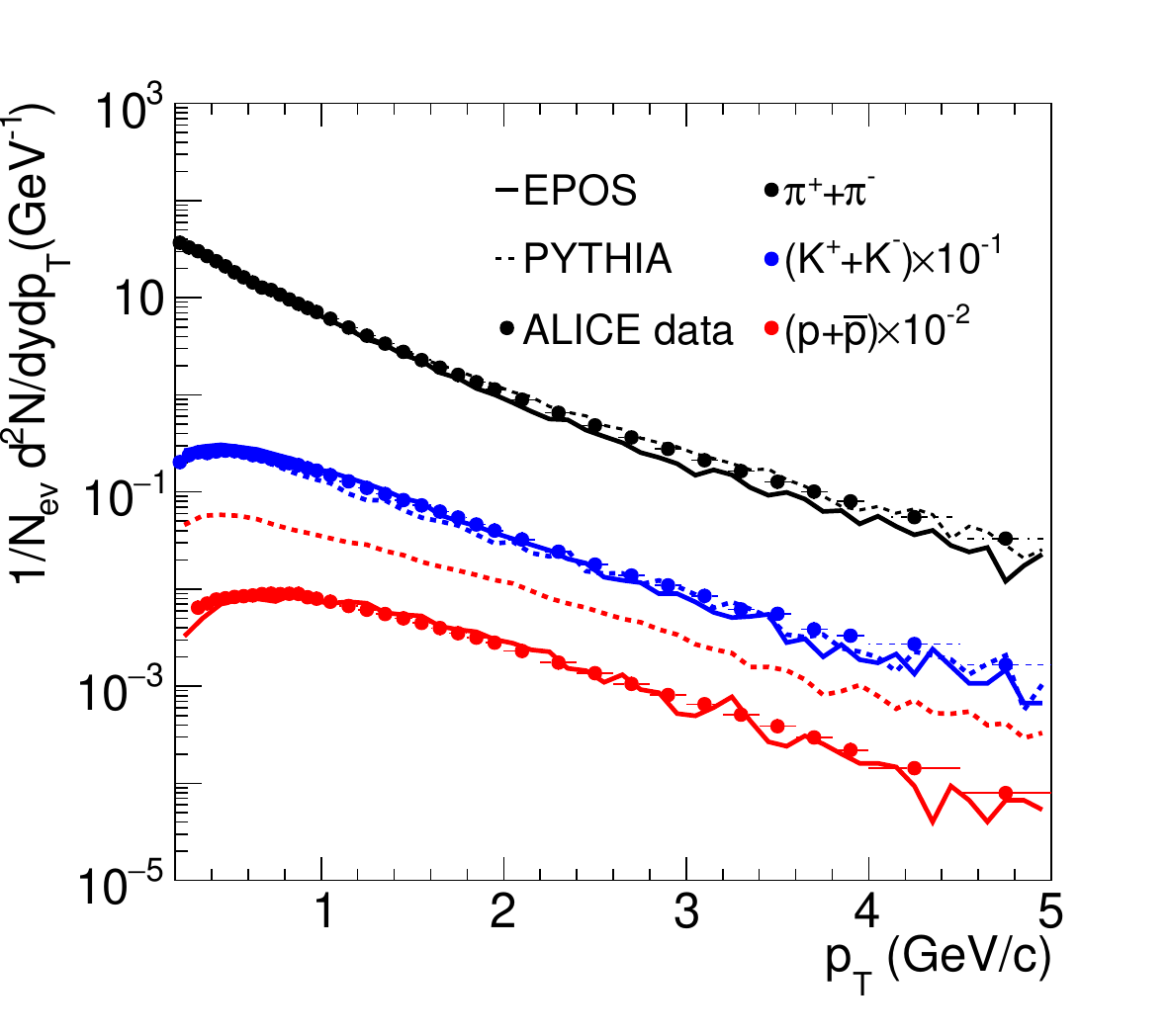}
    \caption{Transverse momentum spectra of $\pi^{\pm}$, $\rm K^{\pm}$, and p+$\rm \overline{p}$ from PYTHIA8 and EPOS4 simulations of pp collisions at $\sqrt{s} = 13.6$~TeV compared with ALICE data from pp collisions at $\sqrt{s} = 13$~TeV~\cite{ALICE:2020nkc} in the 0--1\% multiplicity class.}
    \label{fig:pT}
\end{figure}

We first verified  whether the two event generators produce transverse momentum spectra that approximately reproduce the data measured by the ALICE collaboration in pp collisions at $\sqrt{s}=13$ TeV~\cite{ALICE:2020nkc}. Figure~\ref{fig:pT} displays comparisons between the ALICE data~\cite{ALICE:2020nkc} and the PYTHIA8 and EPOS4 predictions for the 0--1\% cross sections of the production of charged pions ($\pi^{\pm}$), charged kaons ($\rm K^{\pm}$), and protons/anti-protons (p/$\rm \overline{p}$) as a function of their $p_{\rm T}$. One finds that both models describe the experimental data rather well, except for PYTHIA8 which significantly over-predicts the proton and anti-proton production. It is thus legitimate to expect that they might produce a sensible representation of inclusive and species specific balance functions. 

Balance functions and normalized two-particle differential cumulants were computed with Eqs.~(\ref{eq:BF},\ref{eq:R2}) for $\pi^{\pm}$, $\rm K^{\pm}$, and p$\rm \overline{p}$ in the rapidity range $|y| < 1$, the momentum interval $0.2 < p_{\rm T} < 2.0$ GeV/$c$, and over the full azimuth $0\le \varphi < 2\pi$ coverage. The lower $p_{\rm T}=0.2$ GeV/$c$ was selected to mimic the performance achievable by the ALICE detector at the LHC, whereas the upper limit $p_{\rm T}=2.0$ GeV/$c$ was used to limit the impact of jet particles to the strength of the correlation functions. Weak decays of $\rm K^0_{\rm S}$ and $\rm K^0_{\rm L}$ mesons and $\Lambda^0$, $\Sigma$, $\Xi$, and $\Omega$ strange baryons, and their respective anti-particles, were turned off in order to exclusively sample correlations of primary $\pi^{\pm}$, $\rm K^{\pm}$, and p$\rm \overline{p}$ and those produced by decays of strongly-decaying hadron resonances. Leptons resulting from decays of all mesons and baryons were also ignored in the computation of the correlation functions. In order to avoid intricacies associated with error propagation for balance functions, statistical uncertainties were computed using a sub-sample technique based on 10 sub-samples. 

\section{Results}
\label{sec:results}

Balance functions (BFs) were evaluated for primary charged hadrons produced in pp collisions at $\sqrt{s}=13.6$ TeV in the transverse momentum range $0.2 < p_{\rm T}< 2.0$~GeV/$c$ and the rapidity range $|y|<1.0$. Section~\ref{sec:inclusiceBF} presents balance functions of unidentified charged hadrons whereas those of identified $\pi^{\pm}$, $\rm K^{\pm}$, and p$\rm \overline{p}$ are discussed in Sec.~\ref{sec:identified}. The evolution of the integrals of these balance functions with 
the charged particle multiplicity are examined in Sec.~\ref{sec:integrals}. 

\subsection{Inclusive Charged Hadron Balance Functions}
\label{sec:inclusiceBF}

In the context of string based models, such as PYTHIA, the particle production in pp collisions nominally involves  multiparticle interactions in the soft sector and jets in the hard sector. By contrast, in a hydrodynamics model such as EPOS, one might expect an evolving mixture of core and corona features with increasing charged particle multiplicity. It is thus of interest to examine how balance functions 
evolve with increasing charged particle multiplicity as the type and size of jets (in PYTHIA) and the core component (in EPOS) are expected to vary from low to high multiplicity. 

\begin{figure*}[hbt]
    \centering
    \includegraphics[width = 0.3\textwidth,trim={5mm 6mm 19mm 8mm},clip]{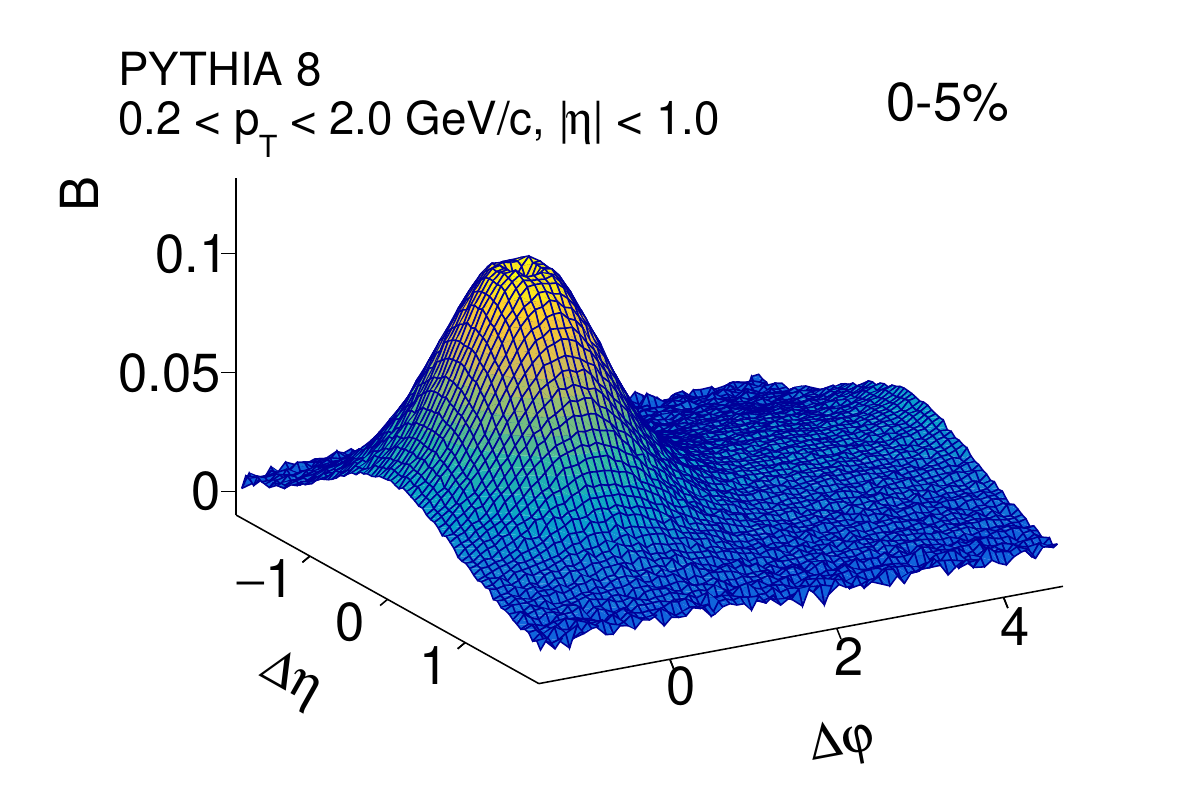}
    \includegraphics[width = 0.3\textwidth,trim={5mm 6mm 19mm 8mm},clip]{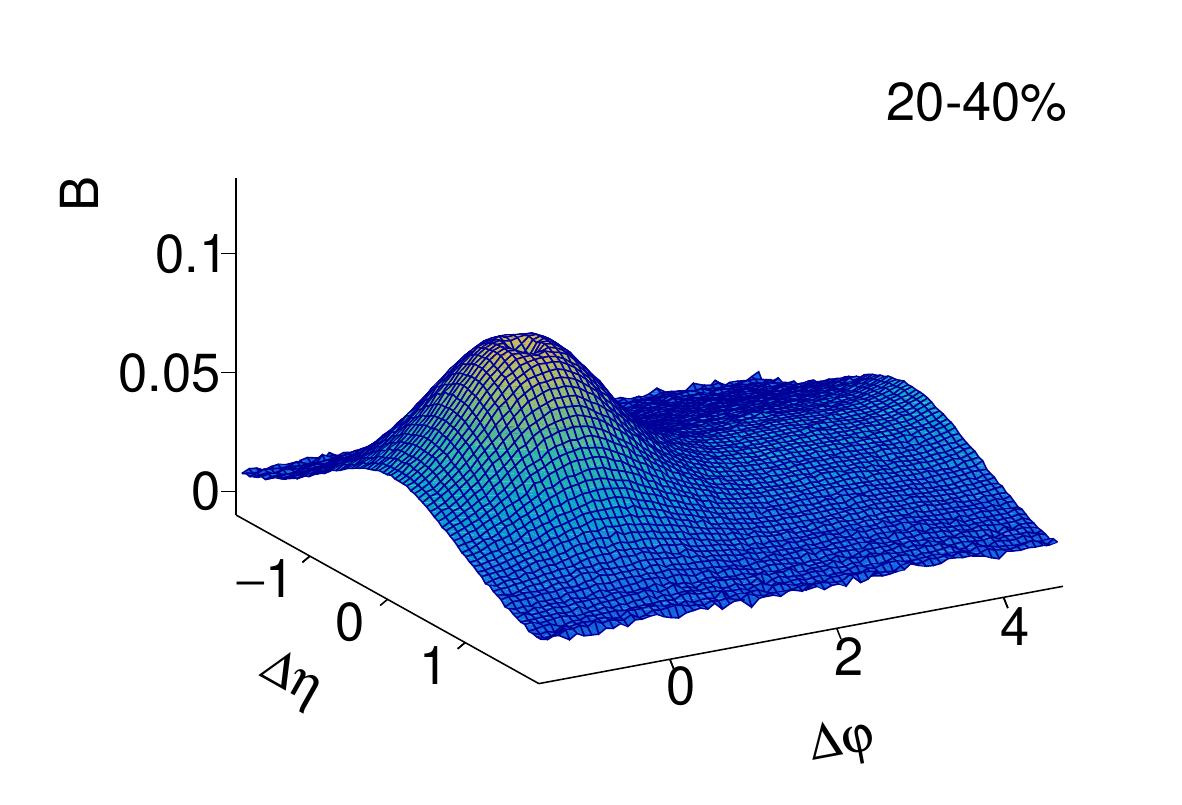}
    \includegraphics[width = 0.3\textwidth,trim={5mm 6mm 19mm 8mm},clip]{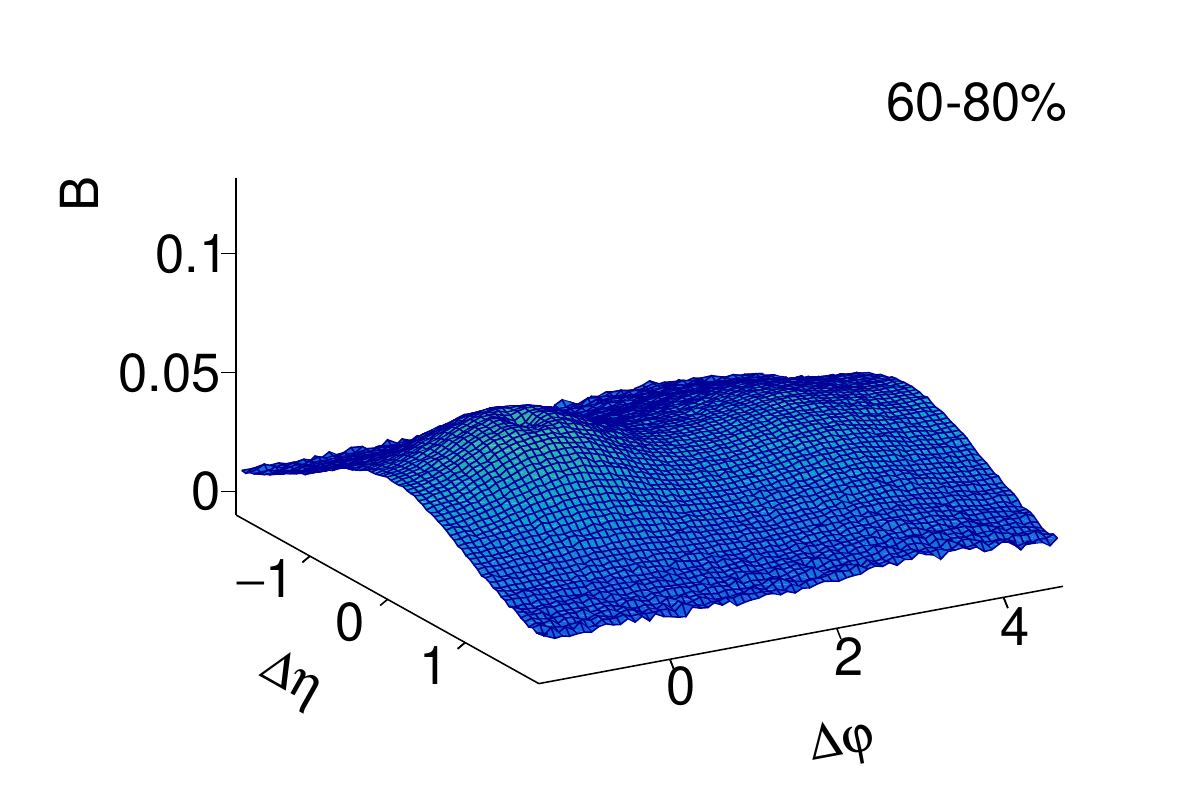}
    \includegraphics[width = 0.3\textwidth,trim={3mm 6mm 19mm 8mm},clip]{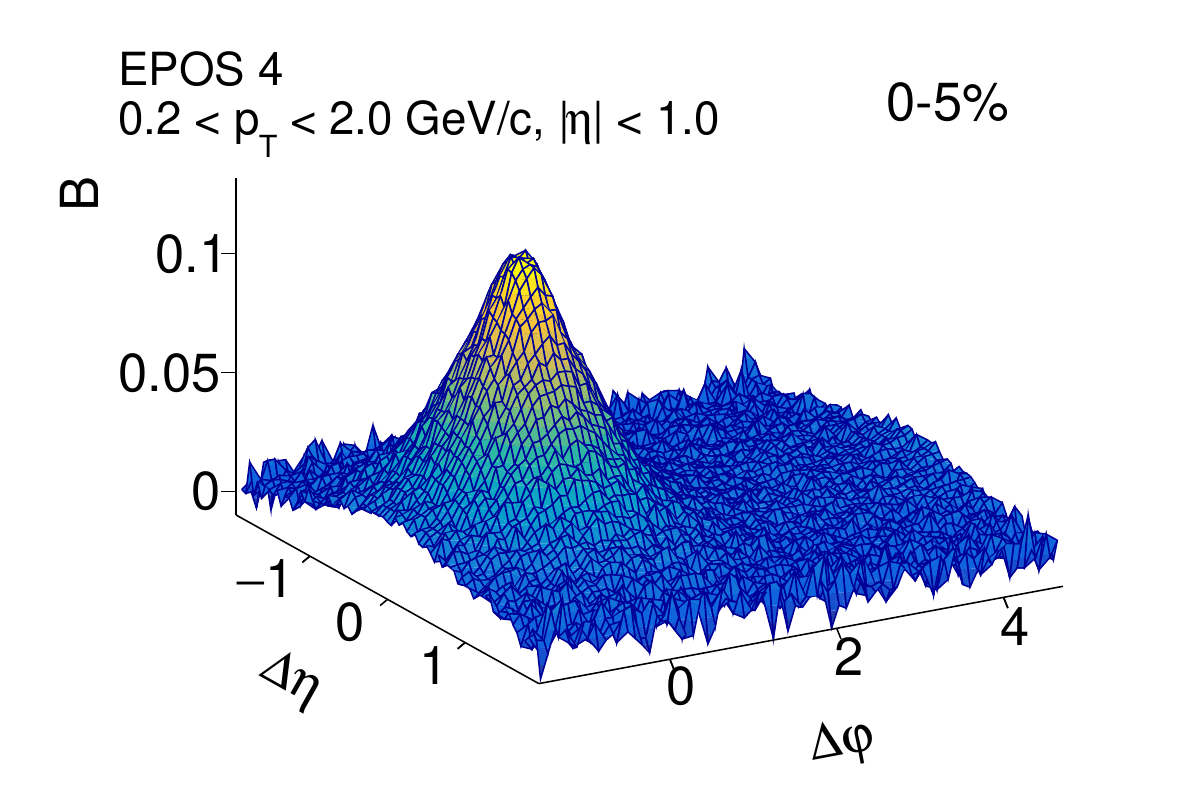}
    \includegraphics[width = 0.3\textwidth,trim={3mm 6mm 19mm 8mm},clip]{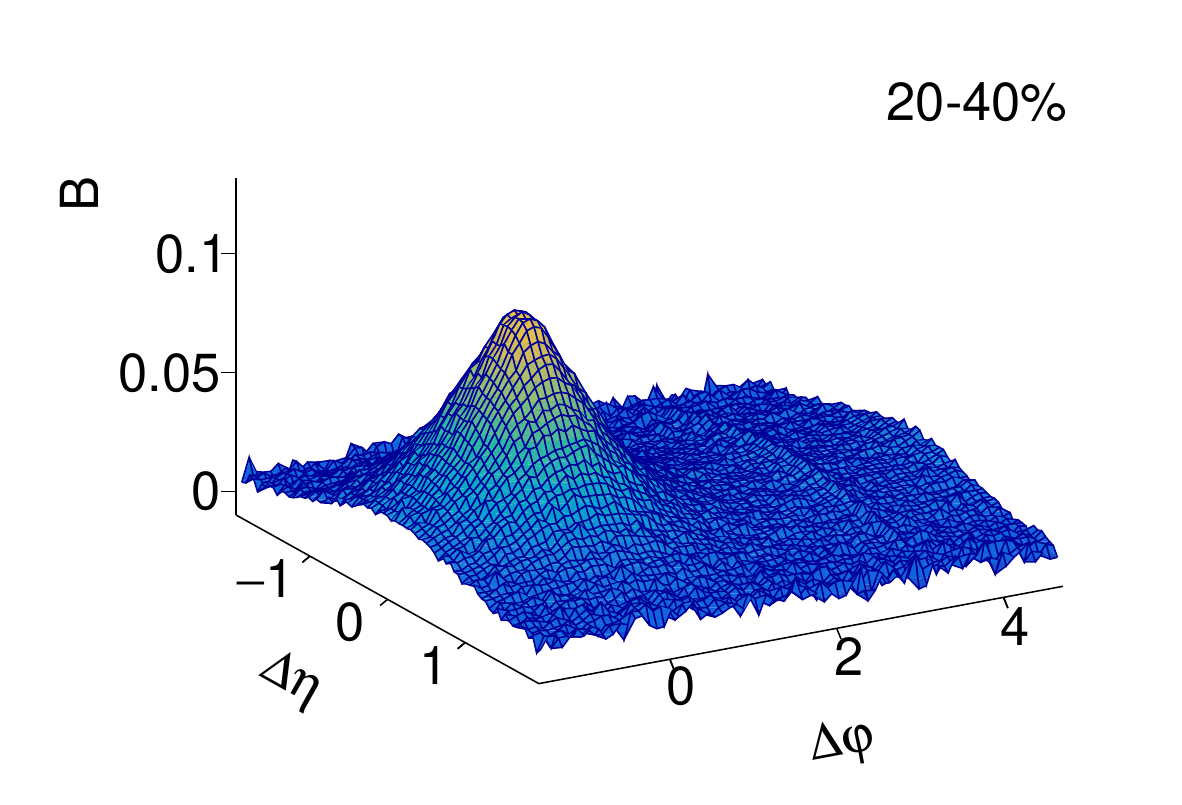}
    \includegraphics[width = 0.3\textwidth,trim={3mm 6mm 19mm 8mm},clip]{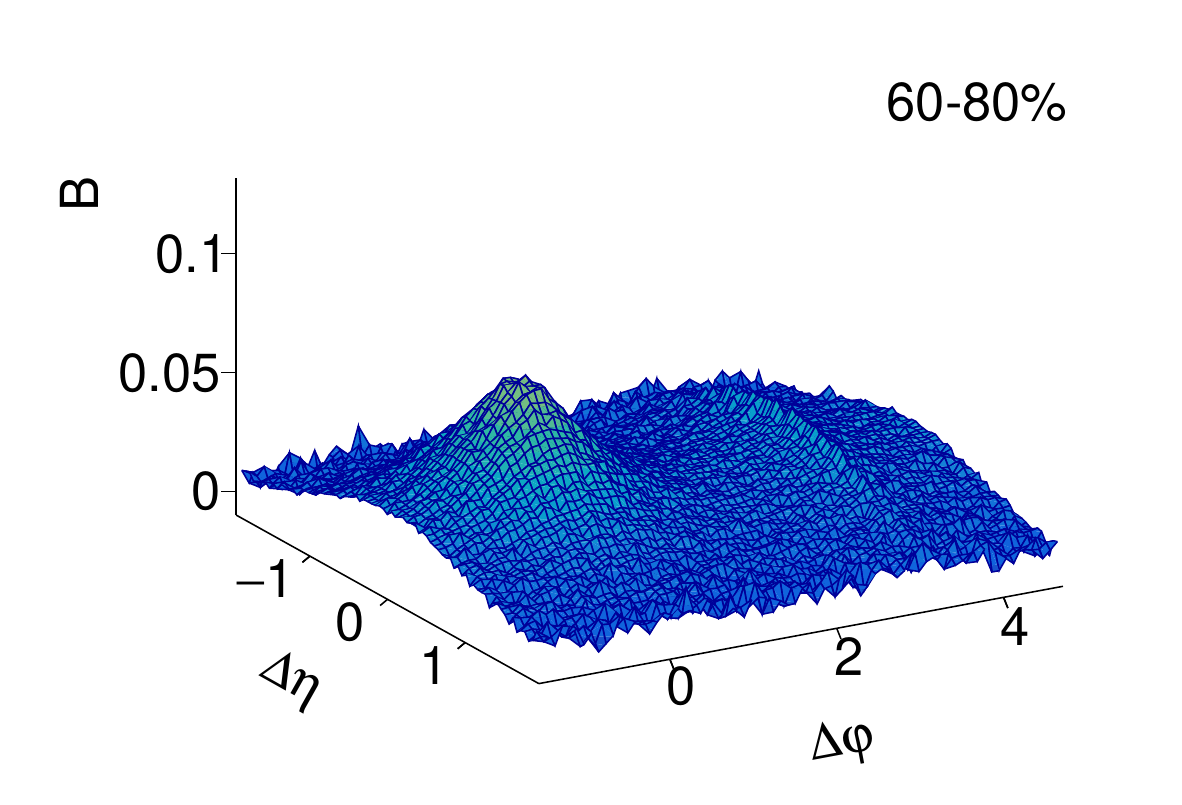}
    \caption{Inclusive charged hadron balance functions measured in the transverse momentum interval $0.2 < p_{\rm T}< 2.0$~GeV/$c$ within $|\eta|< 1.0$ in different multiplicity classes for PYTHIA8 (top) and EPOS4 (bottom).}
    \label{fig:BF_mult}
\end{figure*}

Figure~\ref{fig:BF_mult} displays charged hadron inclusive balance functions computed with PYTHIA8 (top) and EPOS4 (bottom) for collisions corresponding to three selected multiplicity classes.
The BFs from both event generators share qualitative features, including a rather prominent near-side broad peak centered at $\Delta\eta=\Delta\varphi=0$ and a long range, slowly decaying  $\Delta\eta$ component approximately uniform vs. $\Delta\varphi$. One also notices important differences: the near-side of the BF obtained with PYTHIA is somewhat wider and features a shallow dip at its center, whereas the BF obtained with EPOS exhibits a much weaker long-range component and a very narrow ridge-like structure at $\Delta\varphi \sim \pi$ that extends across the full $\Delta\eta$ range of the correlation function. This ridge corresponds to back-to-back emission of hadrons and thus likely origins from  decays of hadron resonances at very small velocity in the collision center-of-momentum frame. The evolution with multiplicity is also slightly different for both generators. The balancing between the near-side and away-side amplitudes with decreasing multiplicity evolves faster in PYTHIA than in EPOS while in the last one the balancing on the away-side is basically focused on the narrow ridge-like structure along $\Delta\varphi \sim \pi$. 

Additionally, one observes that the near-side peak amplitude of BFs obtained with both generators increases with increasing multiplicity while their widths progressively narrow. One also finds the amplitudes of the BF obtained for the different multiplicity classes approximately converge towards one another at the edges of the $\Delta\eta$ acceptance. We thus conclude that BFs obtained with both EPOS4 and PYTHIA8 exhibit narrowing with increasing produced particle multiplicity. While this behavior is expected for isentropic expansion scenarios, as discussed above for hydrodynamic models, and thus potentially anticipated from EPOS4 calculations, it is evidently not expected from PYTHIA8 given this model does not produce a medium endowed with pressure. However, one should also recall that narrowing of correlation functions is known to occur in systems featuring rising average transverse momentum~\cite{Voloshin:2006TRE,Pruneau:2007ua}. Indeed, particles emitted by a common source (e.g., parent particle or cluster) feature an azimuthal and longitudinal separation determined largely by the transverse velocity of the source. The faster the source is, the closer the daughter particles  are expected to be in momentum space, that is, as functions of both $\Delta \eta$ and $\Delta \varphi$. This effect is commonly known as kinematical focusing of the daughter particles. 

\begin{figure}[hbt]
    \centering
    \includegraphics[width = 0.48\textwidth,trim={1mm 1mm 5mm 10mm},clip]{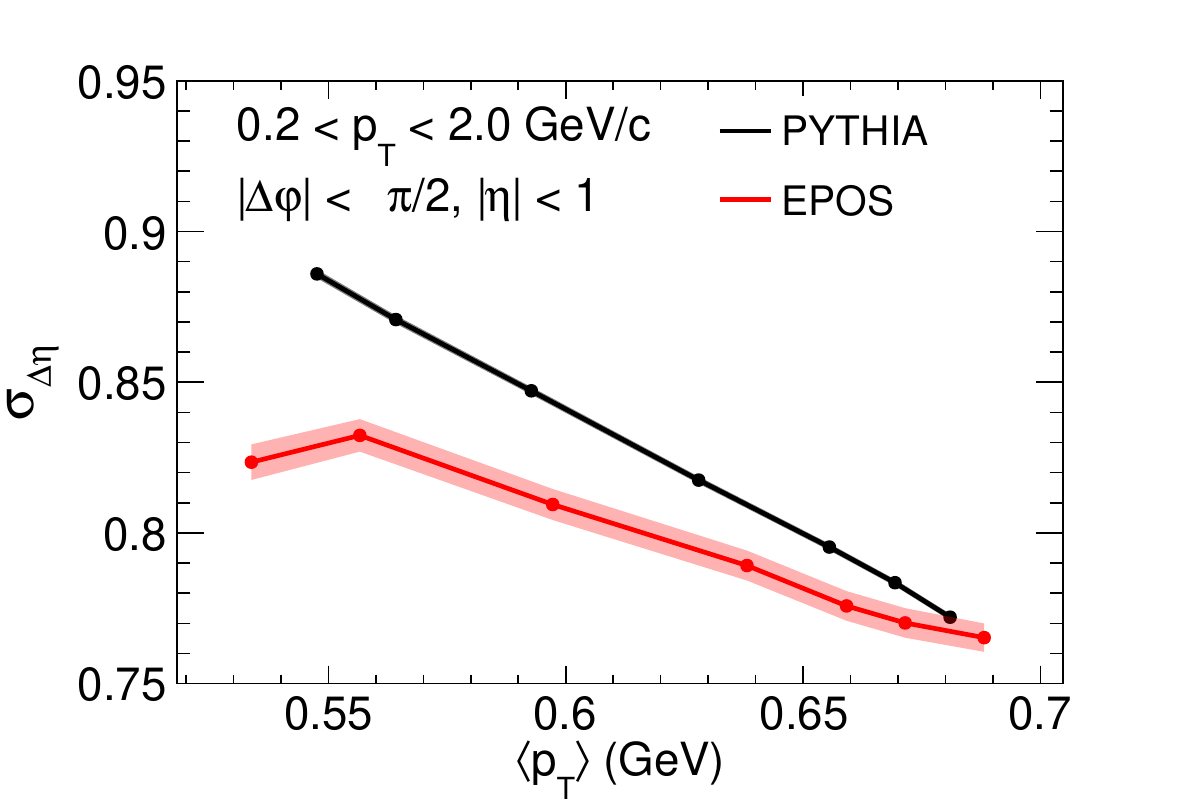}
    \caption{Evolution of the longitudinal rms width of unidentified charged hadron balance functions as a function of mean transverse momentum of each multiplicity class in pp collisions at $\sqrt{s} = 13.6$~TeV for PYTHIA8 and EPOS4.}
    \label{fig:widths}
\end{figure}

We examine effects associated with kinematical focusing in the context of the PYTHIA and EPOS models in Fig.~\ref{fig:widths}, which presents the longitudinal rms widths of the near-side peak of balance functions as a function of the average transverse momentum $\langle p_{\rm T}\rangle$ calculated for each multiplicity class. Charged particles in the range $0.2 < p_{\rm T} < 2.0$~GeV/$c$ and in the same acceptance of $|\eta| < 1$ are used to estimate the mean transverse momentum.  As expected, one finds that the near-side peak width of inclusive particle BFs indeed decreases with rising $\langle p_{\rm T}\rangle$. One notes, however, that the decrease rate of the width vs. $\langle p_{\rm T}\rangle$ is somewhat larger for  PYTHIA balance functions. The difference in widths and rates may result from intrinsic features of the particle production in the two models but it may also reflect, in part, small difference in the relative cross section of pions, kaons, and proton (anti-proton). We thus investigate, in the next section, how BFs of these particles evolve with the multiplicity of the collisions.
%


\subsection{Identified Charged Hadron Balance Functions}
\label{sec:identified}

In the rest frame of a decaying object, daughter particles are emitted back-to-back, with momenta $\vec p\,'_1 = -\vec p\,'_2$, whose magnitude is determined by the difference between the mass of the parent and the sum of the masses of the daughter particles. In the collision center of momentum (CM) frame, the parent object has  a finite  velocity that effectively boosts the daughter particles transversely to momenta $\vec p_1$ and $\vec p_2$. The daughter particles are then kinematically focused towards one another and their angular and longitudinal separations, $\Delta\varphi$ and $\Delta\eta$, reduced in inverse proportion of the velocity of the parent object. A given radial boost nominally has a larger impact on heavier particles which at equal momentum feature a smaller velocity. One then expects that an increase in $\langle p_{\rm T}\rangle$ should have a greater impact on BFs of kaons and protons, relative to those of pions. But the mass difference between parents and daughter particles might also play an important role.

\begin{figure*}[hbt]
    \centering
    \includegraphics[width = 0.3\textwidth,trim={3mm 8mm 16mm 1mm},clip]{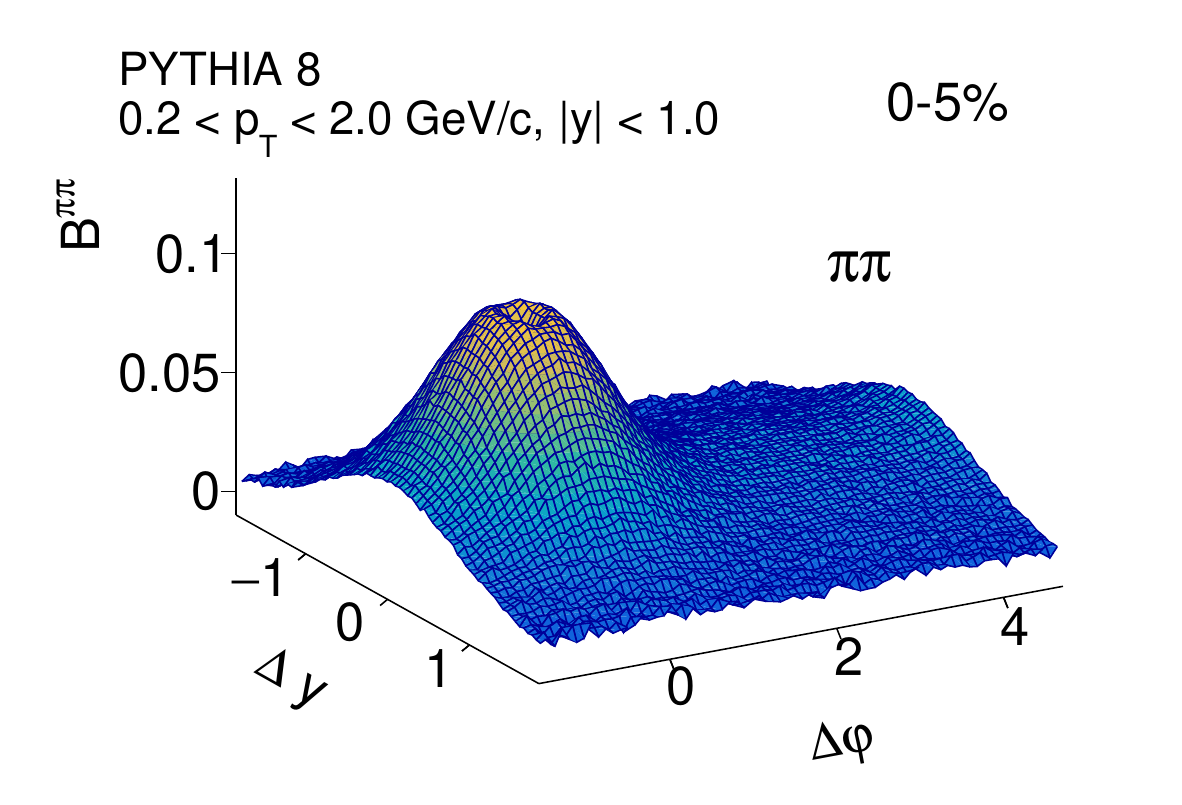}
    \includegraphics[width = 0.3\textwidth,trim={3mm 8mm 16mm 1mm},clip]{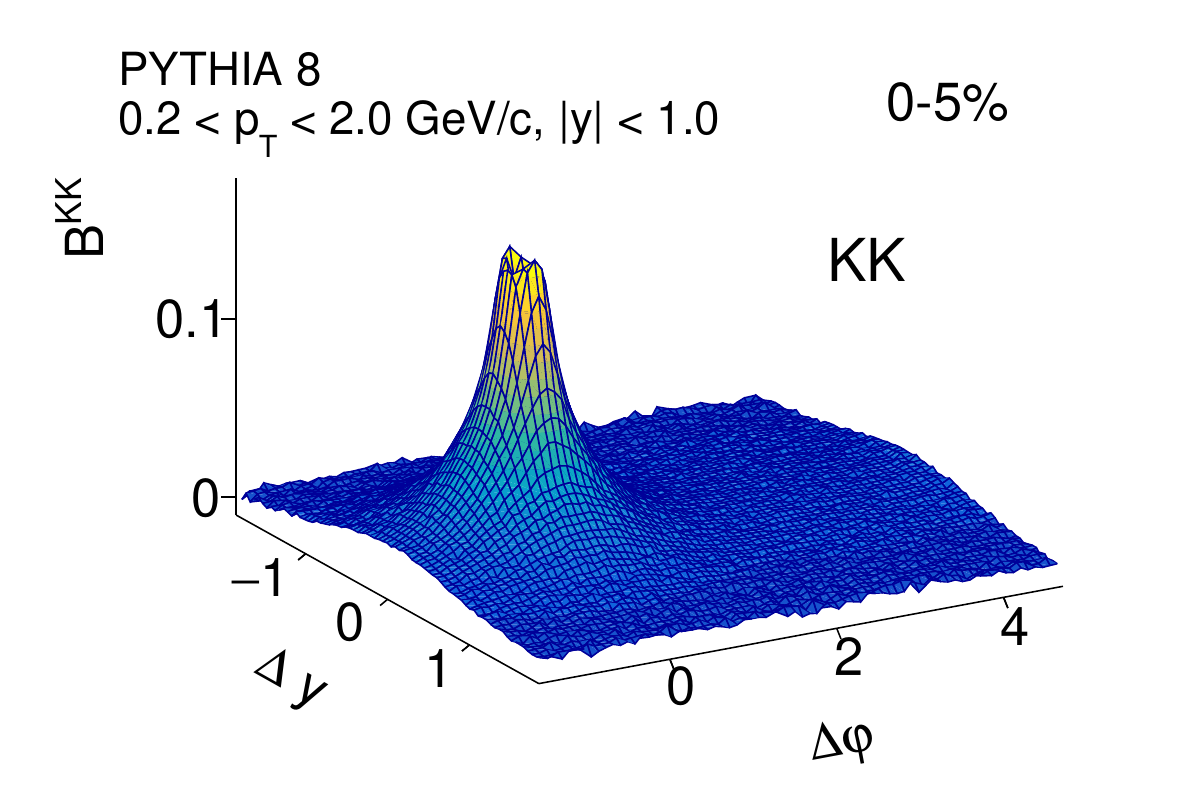}
    \includegraphics[width = 0.3\textwidth,trim={3mm 8mm 16mm 1mm},clip]{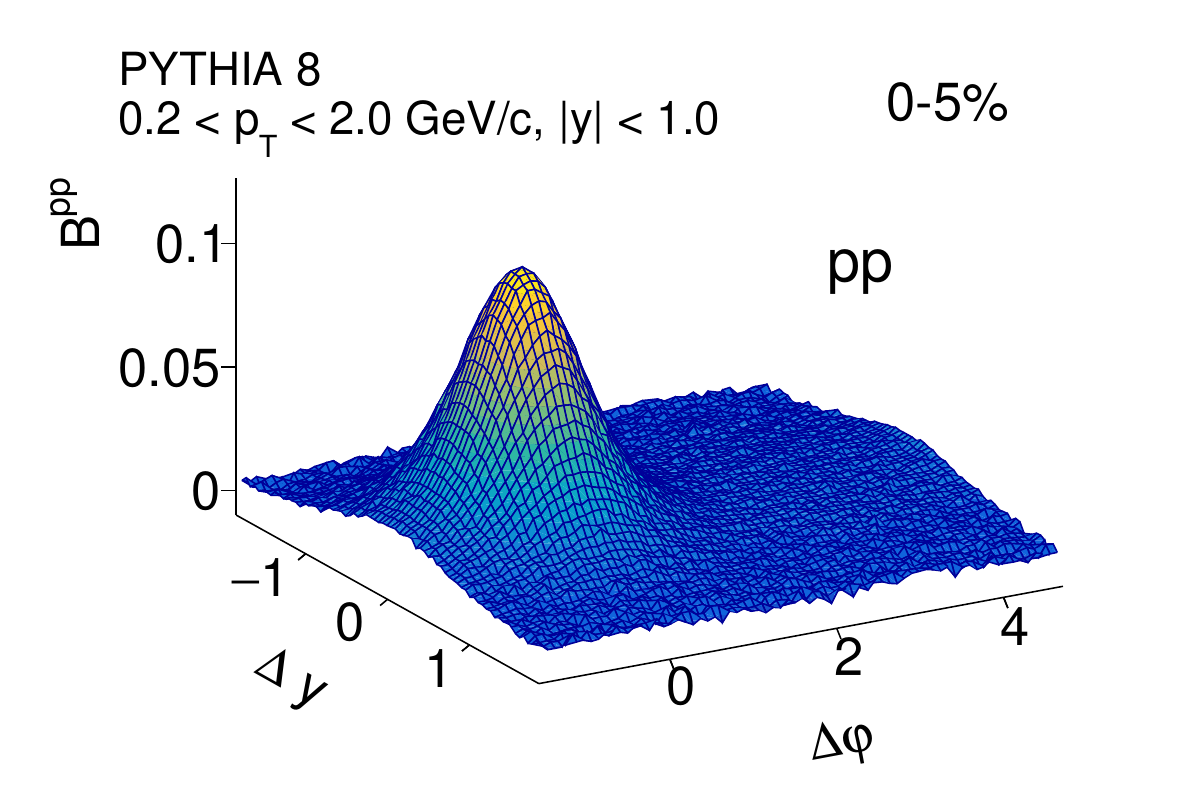}
    \includegraphics[width = 0.3\textwidth,trim={3mm 8mm 16mm 1mm},clip]{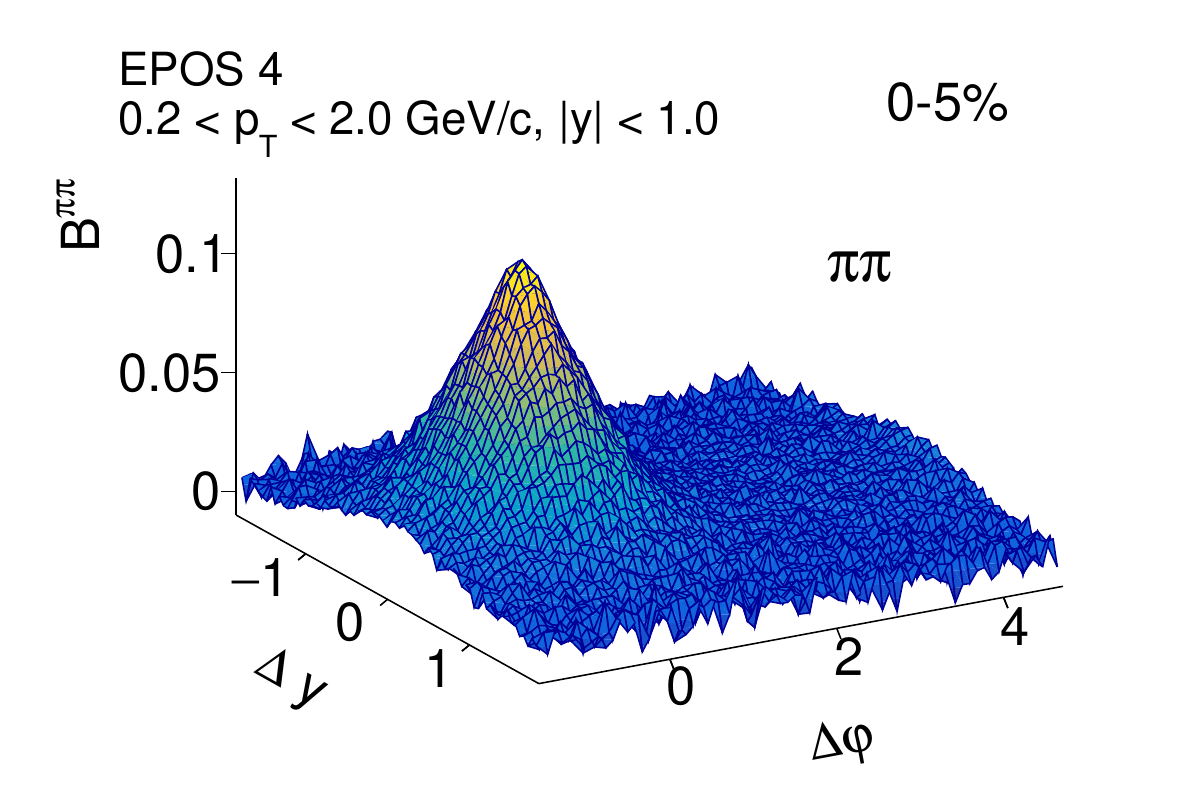}
    \includegraphics[width = 0.3\textwidth,trim={3mm 8mm 16mm 1mm},clip]{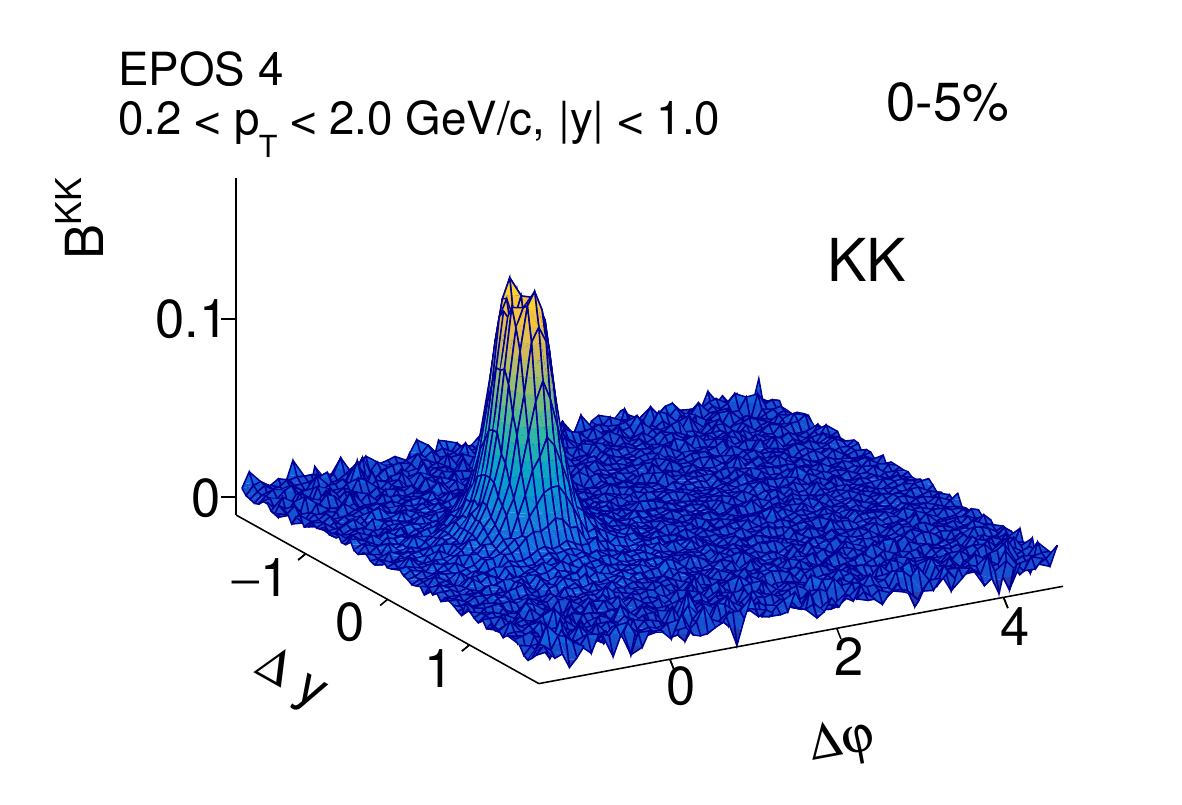}
    \includegraphics[width = 0.3\textwidth,trim={3mm 8mm 16mm 1mm},clip]{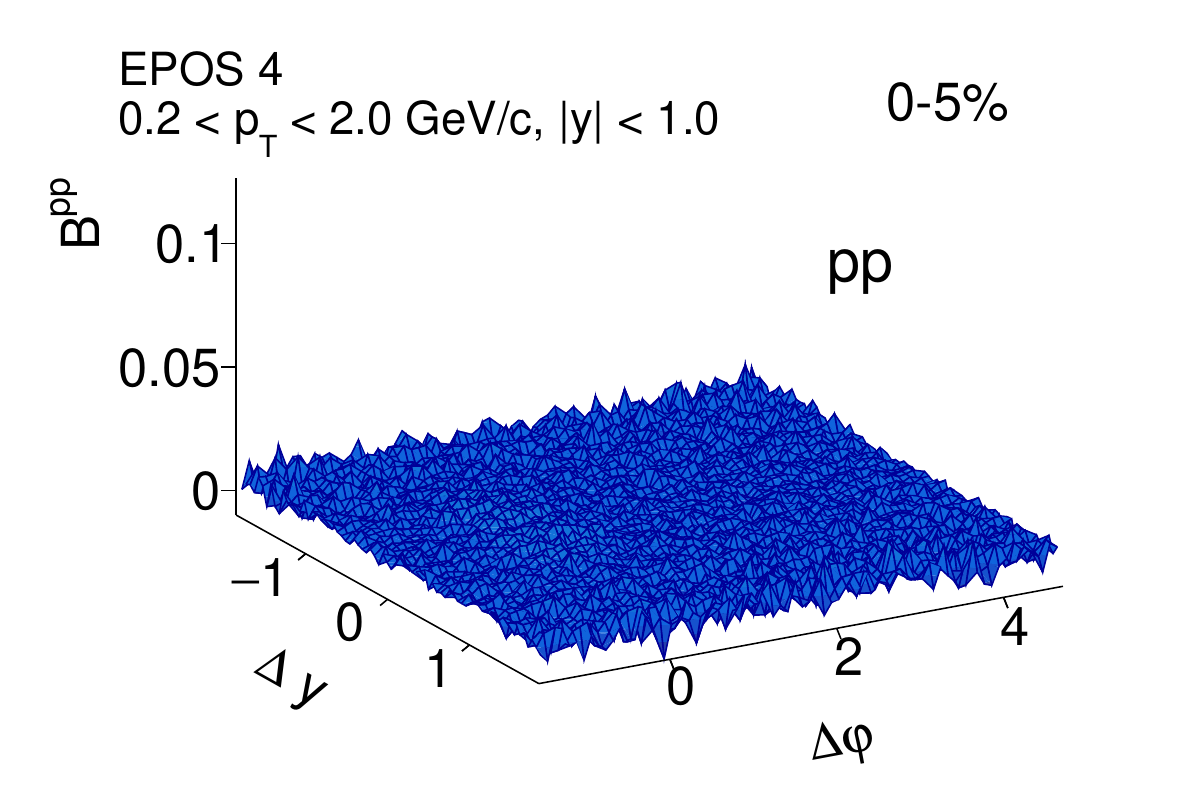}
    \caption{Pion (left), kaon (center), and proton (right) balance functions measured in the transverse momentum interval $0.2 < p_{\rm T}< 2.0$~GeV/$c$ within $|y|< 1.0$ in the 0--5\% multiplicity class from PYTHIA8 (top) and EPOS4 (bottom).}
    \label{fig:identified_BF_mult05}
\end{figure*}

\begin{figure*}[hbt]
    \centering
    \includegraphics[width = 0.3\textwidth,trim={1mm 1mm 16mm 1mm},clip]{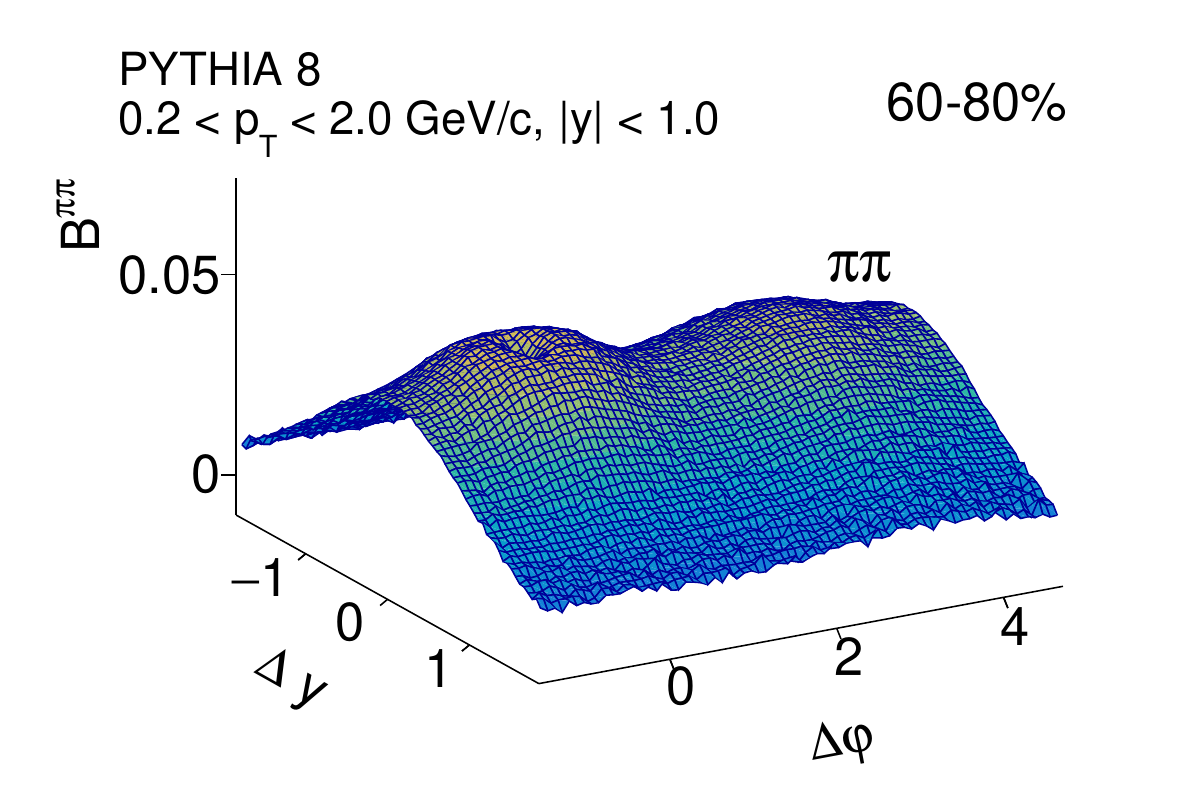}
    \includegraphics[width = 0.3\textwidth,trim={1mm 1mm 16mm 1mm},clip]{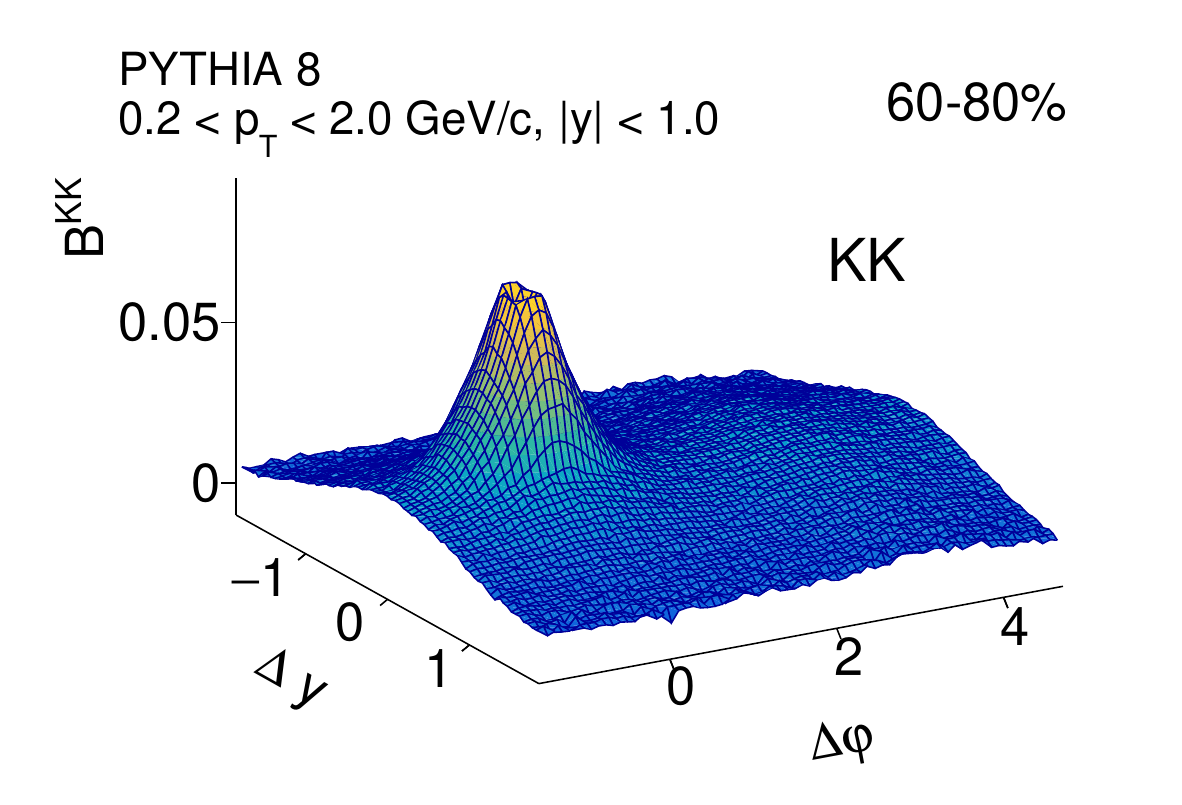}
    \includegraphics[width = 0.3\textwidth,trim={1mm 1mm 16mm 1mm},clip]{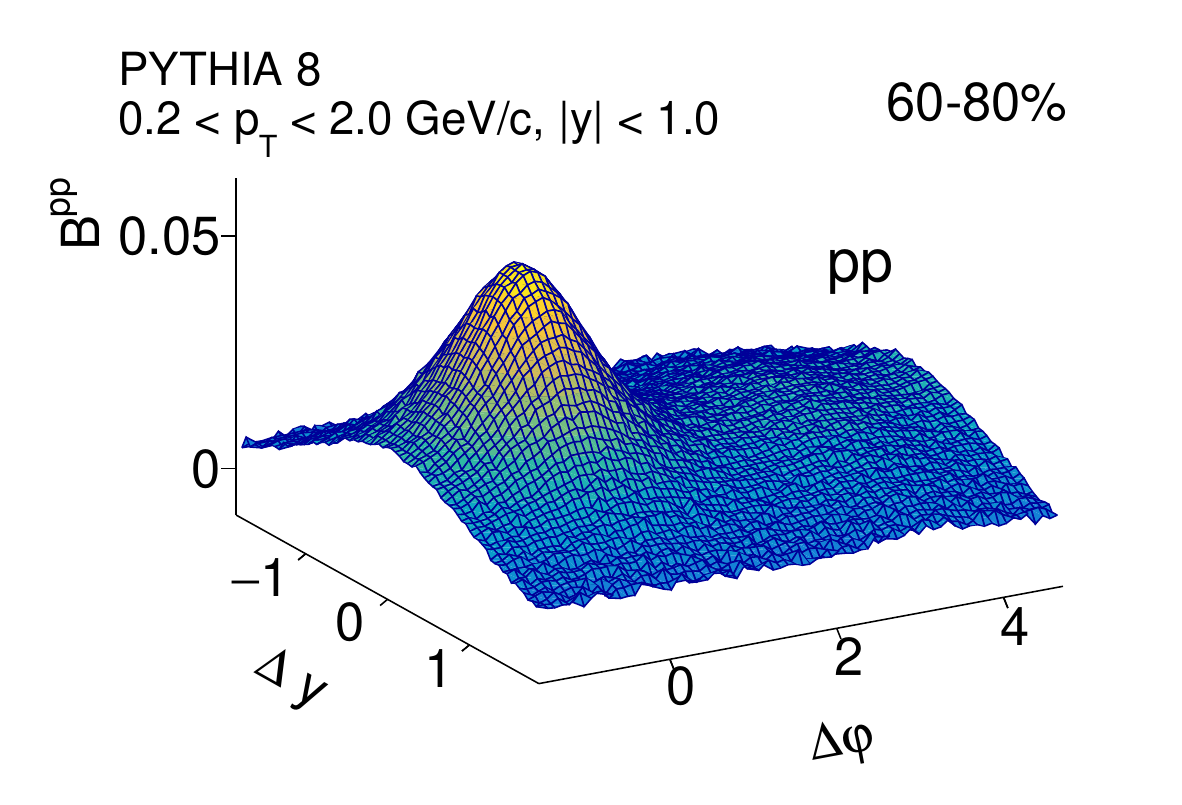}
    \includegraphics[width = 0.3\textwidth,trim={1mm 1mm 16mm 1mm},clip]{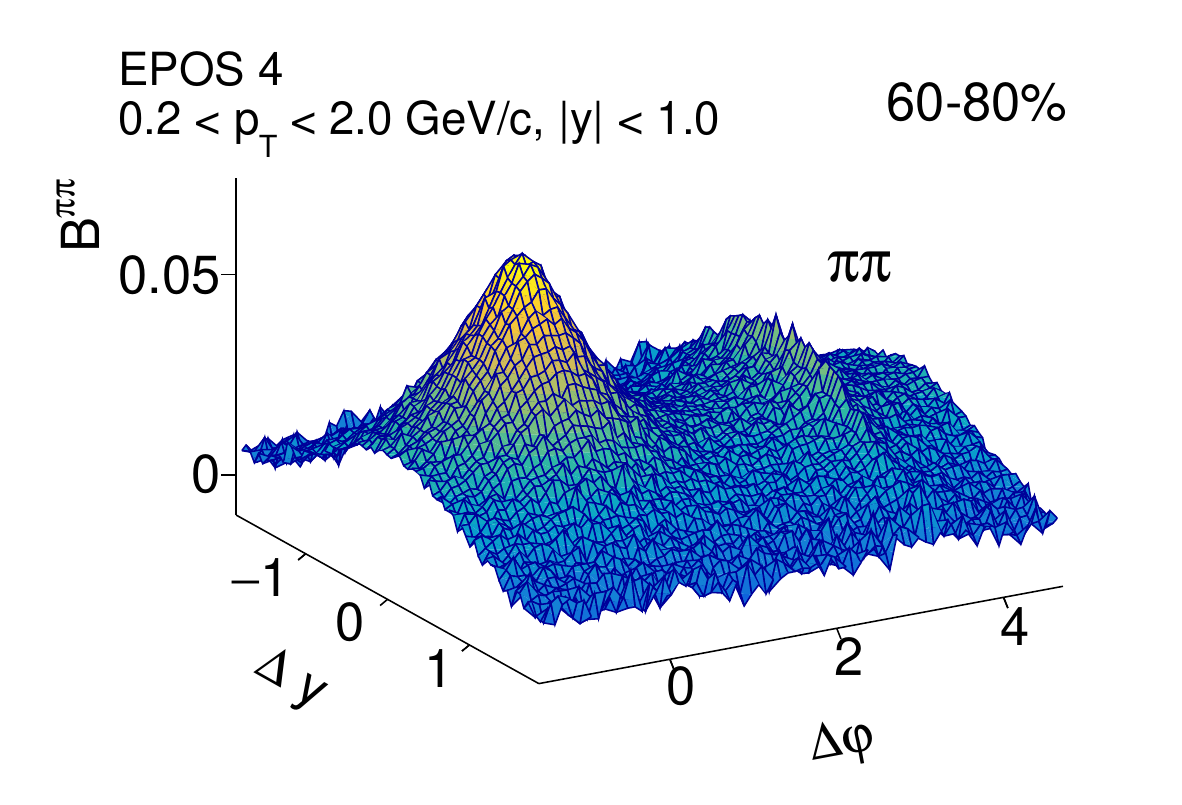}
    \includegraphics[width = 0.3\textwidth,trim={1mm 1mm 16mm 1mm},clip]{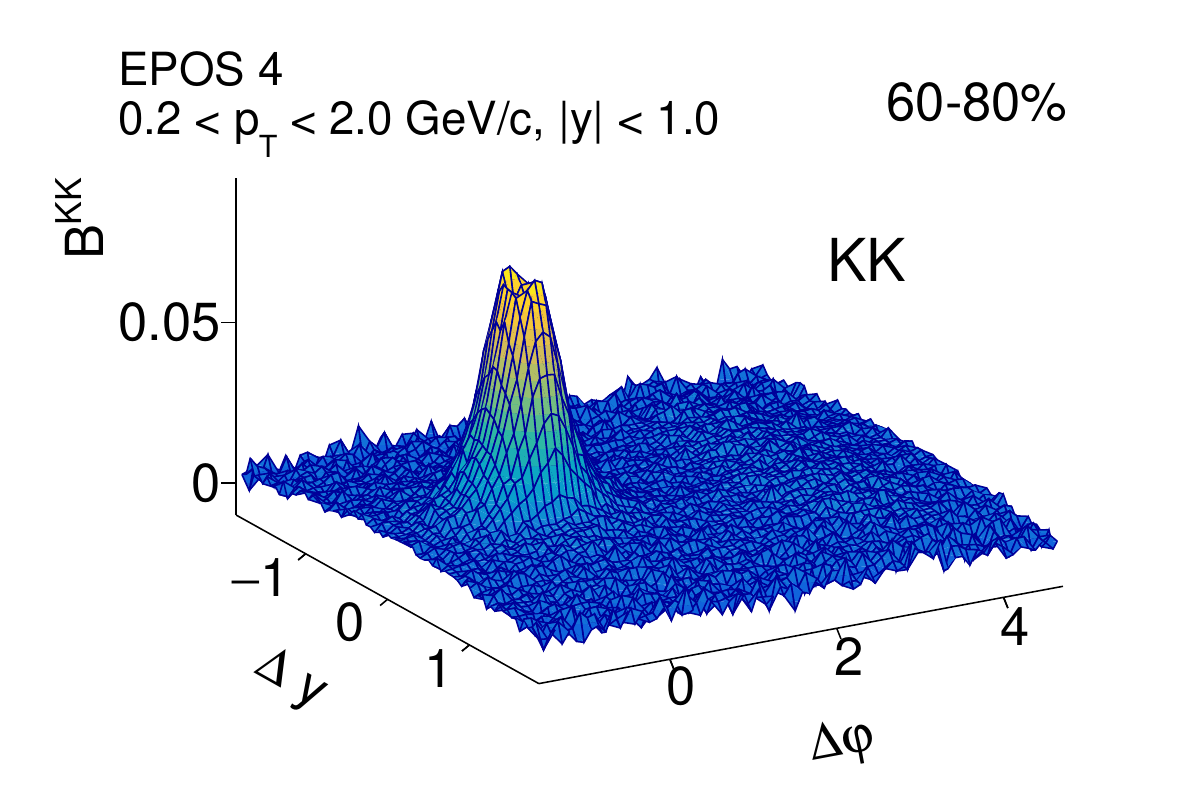}
    \includegraphics[width = 0.3\textwidth,trim={1mm 1mm 16mm 1mm},clip]{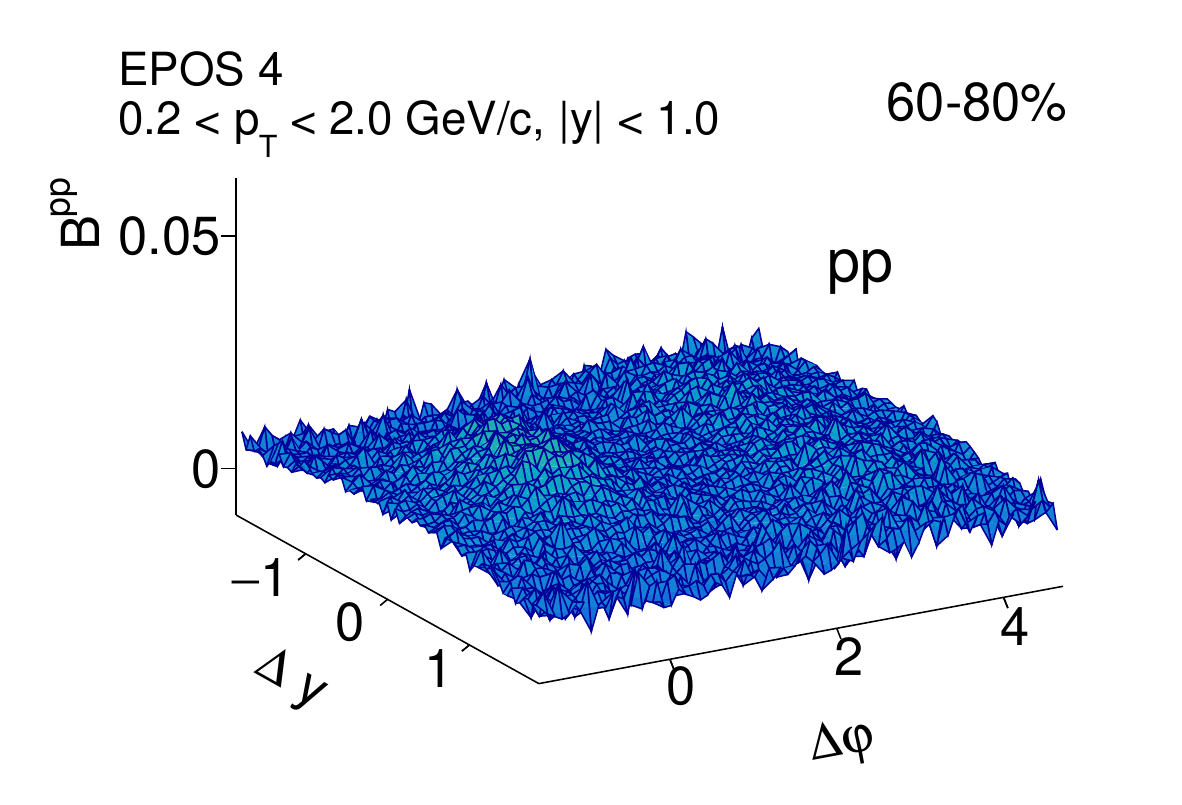}
    \caption{Pion (left), kaon (center), and proton (right) balance functions measured in the transverse momentum interval $0.2 < p_{\rm T}< 2.0$~GeV/$c$ within $|y|< 1.0$ in the 60--80\% multiplicity class from PYTHIA8 (top) and EPOS4 (bottom).}
    \label{fig:identified_BF_mult6080}
\end{figure*}

Figures~\ref{fig:identified_BF_mult05} and~\ref{fig:identified_BF_mult6080} display BFs of pions, kaons, and protons obtained with PYTHIA8 (top) and EPOS4 (bottom) for the 0--5\% and 60--80\% charged hadron multiplicity classes, respectively. Pion BFs obtained with PYTHIA exhibit a dramatic evolution from the lowest to the highest produced particle multiplicity. One finds, in particular, that in low multiplicity events, the pion charge balancing features a very strong away-side component corresponding to approximately back-to-back emission. For increasingly larger multiplicities, however, the charge balancing progressively shifts to the near-side and produces a prominent maximum centered at $\Delta\eta=0$, $\Delta\varphi=0$ in the largest multiplicity class. Additionally, note that the role of the impact of the $\rho^0\rightarrow \pi^+ + \pi^-$ decay is also quite visible vs. an increase in the multiplicity. In the 60--80\% multiplicity class, the near-side features a somewhat broad dip resulting from the large mass of the $\rho^0$-meson relative to the mass of the two pions. But as the multiplicity increases, the transverse momentum also increases and the dip narrows while the magnitude of the near-side peak rises.
The pion BFs obtained with EPOS show a rather different behavior: a prominent near-side peak is already present in the 60--80\% multiplicity class which then exhibit only a modest increase in amplitude in the highest multiplicity events. Also note that the near-side does not feature evidence for the presence of $\rho^0$ decays while the away-side of the BFs features a very narrow ridge centered at $\Delta\varphi=\pi$ across the full $\Delta\eta$ range of the BF.  

The BFs of kaons and proton obtained with PYTHIA and EPOS feature interesting differences relative to the pion BFs and relative to one another. The kaon BF obtained with PYTHIA in the 0--5\% multiplicity class exhibits a tapered near-side peak with a slight depression at its top, which likely originates from $\phi\rightarrow K^++K^-$ decays, and a modest away-side component, while the kaon BF obtained with EPOS features a flat away-side and a rather narrow near-side peak. Interesting differences are also seen in the proton BFs. PYTHIA produces a BF comprised of a modest but finite away-side as well as a prominent and wide near-side peak. It thus seems to favor proton--anti-proton near-side emission as a result of the formation of junctions~\cite{PYTHIA_manual}. In sharp contrast, EPOS produces a proton BF with no away-side yield and a very weak near-side peak. As such, EPOS appears to account for baryon balancing rather differently than PYTHIA. We thus consider, in the next section, the magnitude and evolution of charge balancing by computing the integral of balance functions for inclusive and identified species BFs.

\subsection{Evolution of Balance Function Integrals}
\label{sec:integrals}

By construction, charge BFs obey a simple sum-rule: their integral converges to unity in full acceptance~\cite{Pruneau:2022brh,Pruneau:2023cea}. In a limited acceptance, however, BF integrals amount to a fraction of unity that depends on the BF shape relative to the acceptance of the calculation/measurement. In addition to measurements of differential BFs, it is thus of interest to examine the magnitude of their integral given it reflects the degree to which charge balancing partners are emitted within the acceptance of a particular measurement. It is also of interest to examine whether and how the BF integral evolves with global features of collisions such as the charged hadron multiplicity. Integrals are calculated according to
\begin{equation}
\label{eq:BF_Integral}
    I^{\alpha\beta} = \int_{-\Delta y_0}^{\Delta y_0}  {\rm d}\Delta y  \int_{0}^{2\pi}  {\rm d}\Delta \varphi \,\,   B^{\alpha\beta}(\Delta y,\Delta \varphi),
\end{equation}
where $\Delta y_0$ is the width of the longitudinal  (rapidity/pseudorapidity) acceptance.

\begin{figure}[hbt]
    \centering
    \includegraphics[width = 0.48\textwidth,trim={1mm 7mm 5mm 1mm},clip]{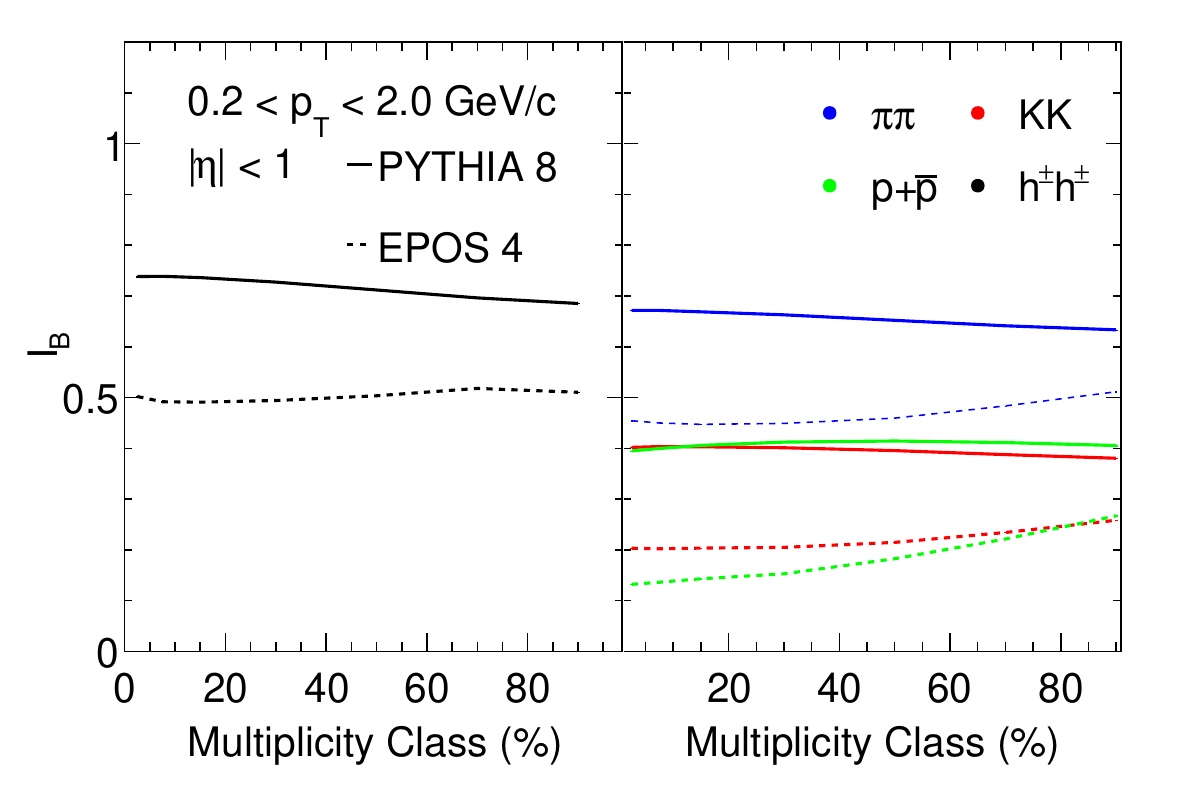}
    \caption{Evolution with multiplicity of the integrals of unidentified charged hadrons (left) and $\rm \pi, K , p\overline{p}$ (right) balance functions in pp collisions at $\sqrt{s} = 13.6$~TeV from PYTHIA8 and EPOS4 event generators. Statistical uncertainties are negligible and thus not shown.}
    \label{fig:integrals}
\end{figure}

Figure~\ref{fig:integrals} (left) displays the evolution of integrals of inclusive BFs computed with PYTHIA and EPOS for selected multiplicity classes. Although the inclusive BFs obtained with EPOS exhibit a significant narrowing with increasing multiplicity, as seen in Fig.~\ref{fig:widths}, their integrals exhibit a small decrease from the 80--100\% to the 0--5\% multiplicity class By contrast, BFs obtained with PYTHIA feature larger integral than those computed with EPOS at all multiplicities. PYTHIA thus ``predicts", overall, that  charge balancing partners are closer in phase space than EPOS does. The integral of PYTHIA BFs additionally exhibits a small, but finite, rise with increasing multiplicity, in contrast to the small decrease observed with EPOS. This thus suggests that the kinematic focusing imparted by the rise in $\langle p_{\rm T}\rangle$ with increasing multiplicity, observed in PYTHIA,  does bring the balancing partners slightly closer in momentum space, also at variance with the behavior with EPOS. Such a minor rise has been reported for charged pion balance functions measured in Pb--Pb collisions by the ALICE collaboration~\cite{ALICE:2021hjb}. However, it is also  interesting to note that the BF integral reported by the ALICE collaboration in pp collisions at $\sqrt{s}=13$ TeV~\cite{baydia} for inclusive charged hadrons is closer to the value produced by EPOS. So while PYTHIA qualitatively predicts the right multiplicity evolution of the integral, it overestimates the integral and EPOS produces, by contrast, an integral compatible with the obtained experimentally but does not feature the observed rise with multiplicity dependence observed experimentally even though it predicts a narrowing of the near side peak of balance functions with rising average transverse momentum.
A more detailed investigation of charge balancing predicted by PYTHIA and EPOS is enabled by considering the integrals of pion, kaon, and proton balance functions shown in the right panel of Fig.~\ref{fig:integrals}. Given that charged pions dominate the inclusive charge production, it is not surprising that the pion BF integrals closely match those of the inclusive BFs obtained by the two models. However, note that the two models predict somewhat different evolutions of the integral  with produced multiplicity. One observes indeed that integrals obtained with EPOS decrease monotonically whereas those obtained with PYTHIA exhibit a small rise with increasing produced multiplicity. The integrals of kaon and proton BFs yield quantitatively distinct (smaller) values but follow qualitatively similar trends to those of the pions. 

Overall, one  observes that EPOS4 and PYTHIA8 predict somewhat different evolution of the BF integrals of identified species. PYTHIA8 predicts a strong focusing of particles resulting in  an increase of BF integrals of both unidentified and identified species. By contrast, while EPOS4 appears to produce narrower BFs in highest multiplicity classes, it predicts a decreasing charge balance of matching partners that appears rather surprising given the decrease of BF widths with increasing produced multiplicity. It is conceivable that this behavior stems from the interplay of corona and core components: as the multiplicity rises, the core component (simulated with hydrodynamics evolution) should progressively weight in. The finite decrease in BF integrals with increasing core contributions may then possibly reflect issues in the handling of the charge balance for particle produced from the core relative to those produced by POMERON exchange (corona). 

\section{Summary and Conclusions}
\label{sec:summary}

This paper presented a comparison of  predictions by the PYTHIA8 and EPOS4 event generators of inclusive and selected species charge balance functions produced in pp collisions at $\sqrt{s} = 13.6$~TeV. PYTHIA simulates particle production based on the LUND string model whereas EPOS uses a combination of core and corona components. The EPOS corona component is based on POMERON exchanges while its core component features hydrodynamic expansion of QGP domains. Although the hydrodynamic component of EPOS does not feature charged or flavor currents, and is thus not strictly suitable towards a study of two-stage quark production and delayed hadronization, it might provide a suitable representation of the evolution of small QGP produced in pp collisions. It might, in particular, provide a reasonable null hypothesis for the presence of two stages of quark production separated by a period of isentropic expansion. Inclusive and identified species charge balance functions obtained with PYTHIA and EPOS share some common qualitative features but disagree significantly on the strength and shape of the correlation functions they predict for both inclusive and identified species charge balance functions. Both models predict the existence of strong near side peaks (corresponding to short range correlations in azimuth and rapidity difference) but the relative magnitude of the near side balancing relative to away-side balancing is found to differ appreciably in the two models. A significant reduction of the longitudinal rms width of the near side peak with multiplicity (and average transverse momentum) is found for both models but PYTHIA predicts a much stronger evolution with the average transverse momentum than EPOS does. The two models also differ in their prediction of the evolution of the BF integrals (of both inclusive and identified species) with produced multiplicity. PYTHIA predicts a modest increase of the BF integrals within the acceptance considered, in line with the predicted significant increase in average transverse momentum, but EPOS yields a slight monotonic reduction of the integrals with increasing multiplicity in spite of the fact that it also features a rise of $\langle p_{\rm T}\rangle$ with rising multiplicity. It is nonetheless interesting to observe that the magnitude of the BF integral predicted by EPOS approximately agrees with recent measurements in pp collisions reported by the ALICE collaboration, even though its evolution with multiplicity differs from observations. The results presented in this work indicate, as expected from general considerations, that the shape and width of balance functions is in large part determined by the underlying physical particle production processes but also underscore the role of increased kinematical focusing associated with a rise in the average momentum of the particles. While EPOS features the expected average transverse momentum increase and associated near side peak width reduction with increased multiplicity, it currently lacks the integration of flavor currents (or quark longitudinal transport) that would enable a more informed discussion of the evolution of the shape and width of balance functions with increasing particle multiplicity. A complete discussion and interpretation of Balance Function observations reported by experiments in pp collision shall thus await further development of EPOS and similar models involving isentropic hydrodynamic system expansion.

\section*{Acknowledgments}

It is a pleasure to acknowledge Klaus Werner for reviewing and providing suggestions to the paper. This work was supported in part by the United States Department of Energy, Office of Nuclear Physics (DOE NP), United States of America, under grant No. DE-FG02-92ER40713 and in part by a grant from the Ministry of Research, Innovation and Digitization, CNCS--UEFISCDI, Romania, project number PN-III-P4-PCE-2021-0390, within PNCDI III. SB acknowledges the support of the Swedish Research Council (VR) and the Knut and Alice Wallenberg Foundation. The authors thank Adrian Sevcenco for the maintenance and operation of the ISS computing grid. 

\bibliography{bibliography}

\end{document}